\documentclass[%
 ,twocolumn%
 ,secnumarabic%
,amssymb, amsmath,nobibnotes, aps, prl, superscriptaddress]{revtex4}
\usepackage{docs}%
\usepackage{bm}%
\usepackage{graphicx}
\usepackage{dcolumn}
\expandafter\ifx\csname package@font\endcsname\relax\else
 \expandafter\expandafter
 \expandafter\usepackage
 \expandafter\expandafter
 \expandafter{\csname package@font\endcsname}%
\fi

\begin{document}

\title{A network criterion for the success of cooperation in an evolutionary prisoner's dilemma, and a variation on Hamilton's rule}%

\author{Stephen Devlin}
\affiliation{Mathematics Department, University of San Francisco, 2130 Fulton Street, San Francisco, CA 94117, USA}
\author{Thomas Treloar}
\affiliation{Mathematics Department, Hillsdale College, 33 E. College St., Hillsdale, MI, 49242, USA}
\date{\today}%

\begin{abstract}
We show that the success of cooperation in an evolutionary prisoner's dilemma on a complex network can be predicted by a simple, quantitative network analysis using mean field parameters.
The criterion is shown to be accurate on a wide variety of networks with degree distributions ranging from regular to Poisson to scale-free. The network analysis uses a parameter that allows for comparisons of networks with both similar, and distinct, topologies. Furthermore, we establish the criterion here as a natural network analogue of Hamilton's classical rule for kin selection, despite arising in an entirely different context. The result is a network-based evolutionary rule for altruism that parallels Hamilton's genetic rule.
\end{abstract}
\maketitle

\def\R{\mathbb{R}}
\newtheorem{dfn}{Definition}[section]
\newtheorem{Not}[dfn]{Notation}
\newtheorem{Theorem}[dfn]{Theorem}
\newtheorem{Lemma}[dfn]{Lemma}
\newtheorem{Proposition}[dfn]{Proposition}
\newtheorem{Corollary}[dfn]{Corollary}
\newtheorem{Def}[dfn]{Definition}
\newtheorem{Remark}[dfn]{Remark}
\newcommand{\leftright}[2]{#1 \hfill #2 \\}
\def\Go{G}
\def\G1{H}
\def\qc{q^c}
\def\qd{q^d}
\def\Gt{G_{t}}


\section{Introduction}

Altruistic behavior among agents in evolving systems, both biological and social, has been widely observed
in nature \cite{Ham,A2,N5,S,H,NS,FF}. The fact that cooperative behavior can emerge between unrelated
individuals in the competitive landscape of natural selection, however, is paradoxical.
Evolutionary game theory is widely employed to address this question, and
models of evolutionary systems that can exhibit realistic phenomena
are of great interest to researchers across disciplines including physics, biology, and the social sciences \cite{A2,N5,S,H,NS,FF,G,Tur,HS,NM,SF,SP,AK,GGC,SRP,DT,HD}.

Complex networks have played a central role in the study of evolutionary systems \cite{NM,SF,SP}. In particular,
network-based models, where evolution is
driven by the discrete replicator dynamics, have been shown to
support robust cooperation that is unsustainable in traditional models built on unstructured populations. In a network-based model, agents occupy the vertices of the network, and interact only within their immediate neighborhood consisting of those agents to whom they are connected by network edges. Interactions take the form of a mathematical game, often the prisoner's dilemma (PD), which captures in a precise framework the temptation to selfishly advance one's own fitness at the expense of a cooperating neighbor. Evolution is implemented using the replicator dynamics.

In \cite{NM}, Nowak and May pioneered the network-based approach by showing that cooperation in the PD could become evolutionarily sustainable on a lattice. In \cite{SP}, Santos and Pacheco further showed that cooperation could even become the dominant population trait on certain {\it heterogeneous} networks like those with scale-free degree distributions, where the network's vertices have degrees that follow an inverse power law. A great deal of subsequent research has focused on the interplay between complex networks and evolutionary dynamics.

In this paper we show that the equilibrium success of cooperation in an evolutionary PD on a network (defined precisely below) can be effectively predicted from a quantitative network parameter derived using a mean field system analysis. Our approach is rooted in the ideas of generating functions associated to random networks \cite{NSW}, and uses results from empirical studies of the dynamics. We compare our theoretical predictions to Monte-Carlo simulations and find excellent agreement across networks with varying topologies and varying average degrees. Given the inherent complexity of these dynamical systems, the accuracy with which the simple criterion here predicts actual dynamics is especially appealing.

Finally, we give an interpretation of the criterion derived here as an analogue of Hamilton's rule for kin selection; a classical result in genetics that explains emerging cooperative behavior among related individuals. In doing so, we are able to apply techniques from the study of complex networks to a modeling problem in evolutionary biology, and arrive at a result that both brings insight to both the factors that drive the model, as well as the social and biological questions that originally inspired the model.

\section{Preliminaries}
\label{network}

The PD is widely studied as a framework in which to model problems involving conflict and cooperation. Two players choose between cooperation (C) and defection (D). Players' strategy choices determine (normalized) payoffs--interpreted as fitness in evolutionary biology--as follows: mutual cooperation gives R to each player, a defector exploiting a cooperator gets T, and an exploited cooperator gets S. Two defectors each give and receive nothing, and payoffs satisfy $T>R>0\ge S$. A rational player will choose to defect since the payoffs for defection strictly dominate those of cooperation regardless of the co-player's strategy.  The result is a Nash equilibrium where both players defect; the dilemma arising from the inefficiency of this equilibrium: both agents would fare better by cooperating \cite{G}.

Define the cost $c$ of cooperation in the PD to be the payoff forgone (from a defector's perspective) by choosing to cooperate, or $c=T-R$. Let $b$ represent the benefit provided to a co-player by a cooperator, so that $b=R-S$. Thus, the cost-to-benefit ratio of the game is
\begin{equation}
\label{cbr}
r=\frac{c}{b}=\frac{T-R}{R-S}.
\end{equation}
The cost-to-benefit ratio indicates of the temptation to defect inherent in the game, with larger $r$ corresponding to a stronger temptation to defect.

A widely adopted payoff normalization \cite{NM,SP,AK,GGC,SRP,DT} sets $R=1$ and $S=0$, so that the game depends on the single parameter $1+r$ indicating the temptation toward defection in the game.  Taking $S$ close to zero amounts to an assumption that social interactions are inexpensive. With this normalization, the game lies on the boundary between the PD and the snowdrift game (SG), another commonly studied game of cooperation. In the SG, the the bottom two PD game payoffs are reversed so that cooperation is a better unilateral response to defection: $T>R>S>P$. In that case, and setting $T=1+r$, $R=1$, and $P=0$, the Nash equilibrium is $1-\frac{r}{S+r}$, so is close to zero as long as $S$ is sufficiently close to zero. Qualitatively, the case of $S=0$ (the so-called weak PD) addresses both games when social interactions are inexpensive, so is the focus from here on. While the $S\approx 0$ assumption is both plausible and widely adopted, it is significant, and dropping it has a considerable effect on system dynamics \cite{Mas}.

Evolution is introduced through repeated interactions between agents with respect to the replicator dynamics. The replicator dynamics model natural selection using agent fitness comparisons that result in stochastic imitation of fitter strategies by less fit strategies (details below). In the repeated PD, payoffs are further required to satisfy $T+S<2R$ in order to ensure that full cooperation in the population remains Pareto optimal.

When a population of agents is unstructured and agent interactions are random, the replicator dynamics favor defection and cooperation is driven to extinction. As mentioned in the introduction, the situation is strikingly different when the population is structured by a network.

Consider a network $\mathcal{N}$ consisting of vertices and undirected edges where neither loops nor multiple edges are allowed.
Agents occupy the network's vertices and are constrained to interact only with their immediate neighbors; those agents with whom they are connected by an edge. A round consists of each agent playing a pure strategy in a PD with all neighbors, and accumulating the resulting payoffs. Following a round of play, agents simultaneously update strategies using discrete replicator dynamics: if agent $x$ has accumulated payoff $P_x$, and compares her payoff to that of agent $y$, then $x$ will adopt the strategy of $y$ with probability
$$P_{x\to y}=\frac{\max\{0,P_y-P_x\}}{(1+r)k_{\text{max}}},$$
with $k_{\text{max}}$ equal to the larger of the degrees of vertices $x$ and $y$ \cite{SP,GGC,SRP,DT,HD}.

We perform simulations on various specific networks (details below) with $10^4$ vertices, in each case starting from a random strategy assignment where the probability of an agent cooperating is $0.50$.
A {\it series} is defined to consist of $10^4$
rounds of play and updating. The {\it series mean} is taken to be the average cooperation level over the last 1000 rounds of the series. For a given network, 100 series are run, and the equilibrium cooperation level is taken to be the average of these 100 series means.

It is well documented, and summarized below, that cooperation can become evolutionarily stable in network models of this kind. Moreover, the extent of the evolutionary success of cooperation has been shown to depend greatly on the particular network topology involved \cite{NM,SP,SF}. In order to explore this phenomenon further, we recall some basic tools in the study of networks.

Let $p_k$ denote the probability that a random vertex from the network $\mathcal{N}$ has degree $k$, and let $X$ be the random variable that takes values in the set of possible degrees of vertices in the network. The probability generating function \cite{NSW} for the distribution of $X$ is given by
$$
G(x) = \sum_{k>0}p_kx^k,
$$
and gives a first-order approximation of network topology. The degree distribution ignores any other contact information present in the network, so $G(x)$ represents a generic network chosen randomly from among all those with the fixed degree distribution. The average vertex degree $V$ in the network is given by $V=G'(1)=\langle k \rangle$.

If an edge is randomly chosen from the network and followed to a vertex at one end, it is $k$ times more likely to lead to a vertex of degree $k$ than a vertex of degree 1. Therefore, if $Y$ is the random variable whose values are the degrees of vertices reached along random edges, then the probability generating function of $Y$ is

\label{uncorrelated neighbor}
\begin{equation}
T(x)=\frac{\sum_{k>0}kp_k x^k}{\sum_{k>0}kp_k} = \frac{1}{G'(1)}\sum_{k>0} kp_k x^k = \frac{xG'(x)}{G'(1)}.
\end{equation}

Define a {\it random neighbor} to be the vertex reached by first choosing a random vertex in the network, followed by a random edge emanating from that vertex. If no degree-degree correlations are present in the network, then it follows that $T(x)$ is the probability generating function for the degree distribution of random neighbors. The average degree of a random neighbor $N$ is therefore the expected value of $Y$, so that $N=T'(1)$. Note that when the probability of an edge leading from a degree $j$ vertex to a degree $k$ vertex is not independent of $j$, then $N$ need not equal $T'(1)$ \cite{DT2}. The mean field parameters $N$ and $V$, of course, require sufficiently large systems to be meaningful \cite{note}.

A critical factor emerging from studies of cooperation phenomena is network heterogeneity \cite{SP,AK,GGC,SRP,DT} . In heterogeneous networks, a broad diversity of vertex degrees are represented. In the context of the evolutionary PD, network heterogeneity has been shown to be strongly correlated with increased success of cooperators \cite{SP,DT}. On certain scale-free networks, for example, cooperation can be the dominant population trait for the full range of PD game parameters.

Heterogeneity can be naturally quantified by the variance of the degree distribution. With $\langle k\rangle$ denoting the expected value of the random variable $X$, and $\langle k^2\rangle$ denoting the expected value of $X^2$, one has $Var[X] = \langle k^2\rangle - \langle k \rangle^2$. Using the notation above,
$$Var[X]= G'(1)T'(1) - G'(1)^2 = V(N-V).$$ If we fix the average network degree $V$, then $(N-V)$, or the size difference between an average neighbor and an average vertex in the network, dictates network heterogeneity.

Since cooperation thrives on heterogeneous networks, $(N-V)$ emerges as a critical network parameter and has been studied in \cite{DT,DT2}. In the following, we introduce a framework that explains this fact in the context of the specifics of the evolutionary PD.

\subsection{Previous Results}

In the evolutionary PD, payoffs flow through connections to cooperators, and agents benefit from maximizing their access thereto.  It is well known that cooperation can thrive through the formation of clusters of cooperating agents \cite{NM,GGC,SRP}. Moreover, when social interactions are inexpensive ($S$ close to zero), cooperators of large degree are especially stable \cite{SP,SRP,DT}. By contrast, as a large defector converts her neighborhood to defectors, she significantly reduces her own payoff and becomes susceptible to takeover by a cooperator. For this reason, larger degree vertices have been shown to disproportionately favor cooperation \cite{SRP}.

A more detailed picture of the dynamics emerges in \cite{GGC}. For low temptation to defect, cooperation is the social norm. As the temptation to defect increases, the dynamics are governed by three populations: a core (or cores) of cooperating agents, a core (or cores) of defecting agents, and a critical fluctuating population of sometime-cooperators and sometime-defectors. The resilience of cooperators, as discussed in \cite{GGC}, is determined by interactions between agents on the border of a cooperator core. When the temptation to defect is too high, defectors eventually invade the core by stripping off layer upon layer of exposed cooperators until they are largely eradicated from the population.

\section{Results and Discussion}

Since the growth or breakdown of a cooperator core is determined by the core's exposure to fluctuating nodes, that is where we focus our analysis. Consider an interaction on the frontier of a cooperating cluster. Using the mean field network parameters $N$ and $V$, we address the question of predicting the particular value of the cost-to-benefit ratio, call it $r_{0.5}$, at which point neither a cooperator nor a defector has an advantage (on average). At $r_{0.50}$ one expects each strategy to be equally successful, and a resulting equilibrium where cooperation and defection are approximately equally prevalent. Thus, $r_{0.50}$ is a predicted threshold at which point the system transitions between dominant (defined as more than $50\%$) cooperation and dominant defection.

Given the description of the dynamics in \cite{GGC} (and outlined above),
we consider interactions between agents on the boundary of a cluster of cooperators with the goal of deriving a criterion to calculate $r_{0.50}$.

Of course, within a cooperator or defector core, no strategy changes occur as
a cooperator (resp. defector) considers the success of a another cooperator (resp. defector).
The dynamics are determined by interactions between differing strategists.

Consider, therefore, a randomly chosen vertex and her randomly chosen neighbor. The two possible configurations are that of a cooperator comparing payoff with a defector and vice-versa.

First we analyze a defecting vertex, representing a chain of potentially advancing defectors, and connected to a cooperating neighbor on the border of a core of cooperators. Our assumptions give the defector, $D_V$, the average network degree $V$, while her cooperating neighbor, $C_N$, has the average network neighbor size $N$.  Finally, assume that $C_N$ has $k$ defecting neighbors while $D_V$ has $l$ defecting neighbors.

\begin{figure}
\includegraphics{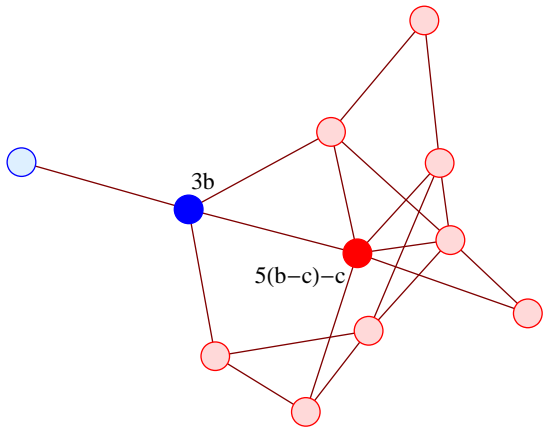}
 \caption{\label{graph_sit}
 The critical frontier between a defector chain (blue) and a cooperator cluster (red). The darker colored vertices are labeled with the respective costs and benefits associated to those strategies and collected over one round of play. Comparison of these values in the context of strategy updating leads to the criterion of Eq. (\ref{normrule}).}
 \end{figure}

We consider the relative costs and benefits associated to the strategies (as opposed to pure payoffs) in
this situation, with costs and benefits as in Eq. (\ref{cbr}). The cooperator perceives the collective value of defection in a round of play to be $(V-l)b$.  The defector, meanwhile, sees the cooperator receiving a benefit $b$ from each of her cooperative neighbors, but also sees this benefit mitigated by the cost of each cooperative act. From the defector's perspective, the value of cooperation is $(N-k)(b-c)-kc$. Fig. \ref{graph_sit} shows the situation when $k=l=1$, that is, when both agents are maximally cooperator connected, and thus have maximal strength in the sense of evolutionary fitness.

Since $r_{0.50}$ is the cost-to-benefit ratio at which point
neither agent perceives an advantage in the other's strategy, we can predict $r_{0.50}$ by equating the collective benefits of the $D$ strategy and the $C$-strategy. This gives
$$
(V-l)b = (N-k)(b-c) - kc.
$$
Rearranging, we get
\begin{equation}
\label{rule}
\frac{c}{b} + \frac{(k-l)}{N} =\frac{N-V}{N}.
\end{equation}
Substituting $c=(T-R)$ and $b=(R-S)$, and inserting the normalized payoffs $T=1+r$, $R=1$, and $S=0$ gives
\begin{equation}
\label{normrule_nbr}
r + \frac{k-l}{N}  =\frac{N-V}{N}.
\end{equation}

Finally, we add an assumption that near $r_{.50}$, the term $\frac{k-l}{N}\approx 0$.
 This amounts to an assumption that when an average cooperator and average defector are connected by an edge, their respective numbers of defecting neighbors are comparable. That this assumption is true in actual simulations is verified below.

With the added assumption the condition simplifies to
\begin{equation}
\label{simpnormrule}
r_{.50}=\frac{N-V}{N}.
\end{equation}

Defectors will advance against cooperators when the equality becomes an
inequality reflecting an inherent advantage to defectors on the frontier.

Conversely, consider a randomly chosen cooperator $C_V $ on a frontier, compared to her randomly chosen neighbor $N_D$.
As above, let $l$ and $k$ be the numbers of defector neighbors of $C_V$ and $N_D$, respectively. The same argument this time yields the condition
$$
r = \frac{V-N}{V} + \frac{l-k}{V}.
$$
 Now, however, the condition is statistically unsustainable since $(V-N) \le 0$, and $\frac{c}{b}>0$. During the system's evolution, smaller cooperators not inside a cluster are stripped away.
As a result, cooperators must migrate to larger vertices in order to survive, leaving the first scenario already analyzed above.

This framework suggests the following: natural selection favors cooperation when the cost-to-benefit ratio of the cooperative act is smaller than the relative size difference between an average neighbor and an average vertex, divided by the average neighbor size:
\begin{equation}
\label{normrule}
r<\frac{N-V}{N}.
\end{equation}
The term $\frac{N-V}{N}$, which is purely a network parameter as well as a natural measure of network heterogeneity on a zero-to-one scale, should mark the phase transition from dominant cooperation (more than half the population cooperating) to dominant defection (more than half the population defecting).

\begin{figure*}
\includegraphics[width=8in]{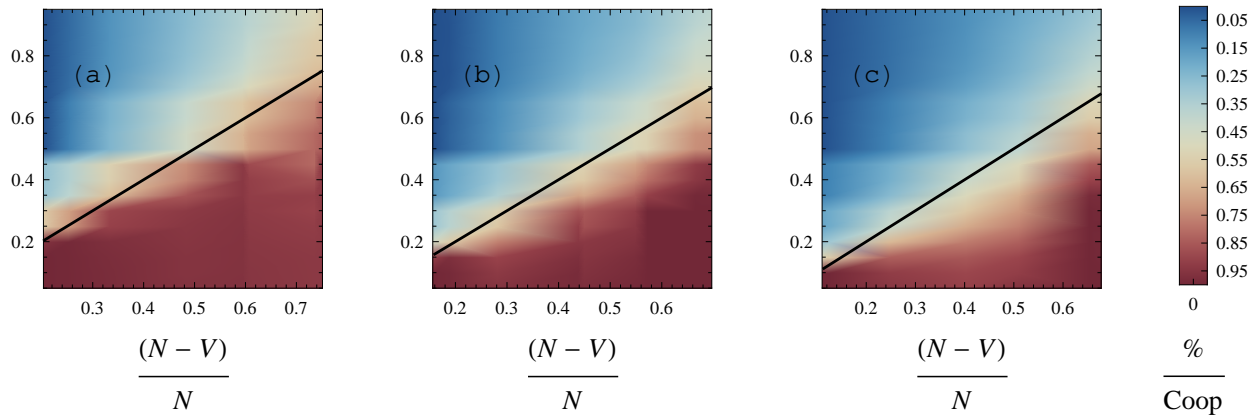}
 \caption{\label{fogs} Simulation results for the evolutionary PD on networks with varying heterogeneity, and average degree 4 (a), 6 (b), and 8(c). The equilibrium level of cooperation is given as a temperature plot depending on both the network parameter $\frac{N-V}{N}$, and the cost-to-benefit ratio $r$. A point $(x,y)$ in the plot, therefore, is colored according to the equilibrium cooperation level in the evolutionary game on a fixed network with heterogeneity given by $x=\frac{N-V}{N}$ and PD cost-to-benefit ratio $y=r$.}
 \end{figure*}

\begin{figure}
\includegraphics{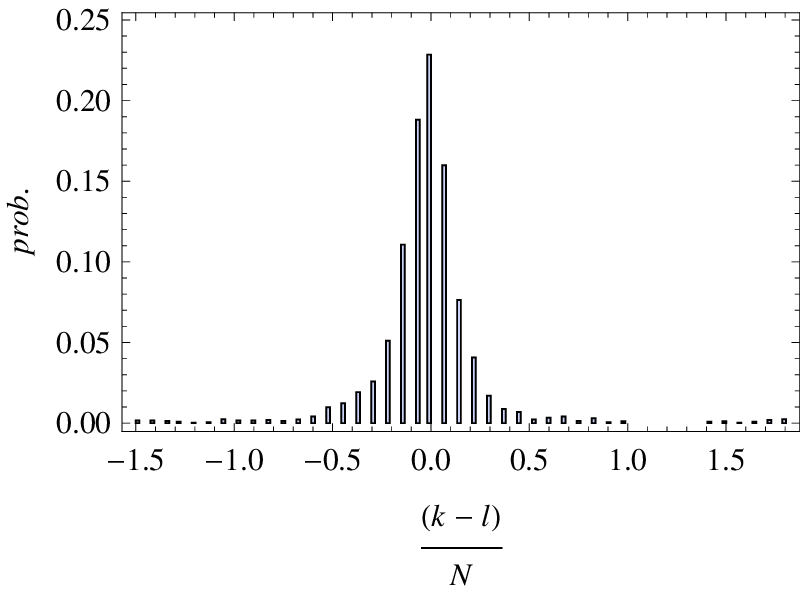}
 \caption{\label{hist} The actual probability distribution of $\frac{(k-l)}{N}$, the difference in the number of $D$-neighbors over all $C$-$D$ edges. Simulations were run on a network with $5\times 10^3$ vertices, and with average degree $V=6$ and average neighbor degree $N=13.35$. The data in the histogram was collected from simulations performed with $r=0.50$, and equilibrium cooperation level of $0.67$ after $10^4$ rounds of play. At $r=0.55$, cooperation levels are below $0.50$.}
 \end{figure}

To test the accuracy of Eq. (\ref{normrule}), we perform simulations on networks with various degree distributions. Networks with $K$ vertices, and average degree $2m$ are constructed via a two step process introduced in \cite{GGM}. First, a network is generated using the algorithm in \cite{GGM}. This algorithm uses a single parameter $\alpha$ to interpolate between an Erd\H{o}s-R\'{e}nyi random network (ER) \cite{ER}, and a Barab\'{a}si-Albert scale free network (BA) \cite{AB1}. Starting from a complete graph on $n_0$ vertices, one of the remaining $K-n_0$ vertices is chosen. This vertex has $m$ edges to attach as follows. With probability $1-\alpha$, the vertex attaches an edge to an existing vertex with a probability proportional to the existing vertex's degree (i.e., by preferential attachment). With probability $\alpha$, the edge is connected to any of the existing $K-1$ network vertices with a fixed probability. This procedure is repeated $m$ times, once for each edge. When $\alpha=0$, one obtains a BA network with a power law degree distribution, and when $\alpha=1$ one obtains an ER random graph. Intermediate $\alpha$ give hybrid distributions with intermediate levels of heterogeneity between the heterogeneous BA networks and the essentially homogeneous ER random networks. Networks are generated with $K=10^4$ vertices, and average degree $2m\in\{4,6,8\}$. For each value of $2m$, networks are generated with $\alpha\in\{0.00, 0.10, 0.20,0.40, 0.60, 0.80, 1.00\}$. Finally, each network is distilled down to its degree distribution by throwing away all other contact information, and a new uncorrelated network is reconfigured, consistent with that degree distribution, using the configuration model \cite{MR}. The result is a maximally random network with the specified degree distribution.

The point of this choice of networks is not any particular topology, but rather, to give a range of varied topologies with varied heterogeneity as measured by $\frac{N-V}{N}$. The results of the simulations described above are consistent with previous work, and are summarized in Fig. \ref{fogs} below.

The theoretical prediction of equation \ref{normrule} is compared with actual simulated dynamics on the networks above in Fig. \ref{fogs}.

The temperature plot in Fig. \ref{fogs} shows that the data are in excellent agreement with the theoretical predictions, proving the effectiveness of the mean field parameters $N$ and $V$, and validating the framework leading to Eq. (\ref{normrule}).

In particular, the black lines in each panel mark the predicted (by Eq. \ref{normrule}) $r_{.50}$, as a function of $\frac{N-V}{N}$, where cooperators and defectors are expected to be at equal strength, and so, are predicted to be equally prevalent. Actual transitions from dominant cooperation (darker red) to dominant defection (darker blue) occur in the neutral tan colored regions between red and blue. Indeed, the black prediction line passes through the neutral, or nearly neutral, regions of the temperature plot.

Statistical fluctuations are most prominent when $V=4$, while for $V=6$ and $V=8$, the predictions are extremely accurate. This is not surprising in light of Eq. (\ref{normrule_nbr}), and the fact that when $V=4$, values of $N$ are smallest, and the term $\frac{k-l}{N}$ most affects the value of $r_{.50}$.

We note that the mean field framework leading to Eq. (\ref{normrule}) is extremely versatile, giving accurate predictions across networks with very different distributions, different levels of heterogeneity, and different average degrees. Additionally, we have checked that the criterion also holds on random regular graphs with average degrees $4$, $6$, and $8$, where cooperation is virtually eliminated immediately, as predicted by Eq. (\ref{normrule}).

We turn to the assumption that simplifies Eq. (\ref{normrule_nbr}); namely, that the numbers of $D$-neighbors of the average agents at either end of a $C$-$D$ edge are comparable. While the accuracy of the predictions in Fig. \ref{fogs} serves as a partial justification, we can consider the assumption directly in the simulations. Fig. \ref{hist} shows a histogram of the distribution of differences in the numbers of $D$-neighbors over all edges connecting a cooperator and a defector, normalized by $N$. The data was collected after the ten-thousandth round of play and updating on a network with $10^4$ vertices, average degree $V=6$, average neighbor degree $N=13.35$, and for $r=0.50$. Note that $\frac{N-V}{N} = 0.55$, at which point cooperation levels are below $50\% $. The distribution of $\frac{k-l}{N}$ is sharply peaked at $0$.  We then let the system run for another $10^3$ rounds, computing the average value of $\frac{k-l}{N}$ after each round. Finally, averaging over these $10^3$ data points gives an overall average value of $\frac{k-l}{N}=-0.046$.

There is also a nice connection between the criterion of Eq. (\ref{normrule}) and the results in \cite{DT}. In that paper the  authors showed that the weighted (by the cost-to-benefir ratio $r$) average equilibrium cooperation level--call it $y$-- on the network depended on the network parameter $x=\frac{V}{N}$ in a linear way. The regression line in that paper gave was given by $y=-1.0074x+.9322$. Notice that the regression line is very close to $y=-x+1$. Inserting $x=\frac{V}{N}$, one gets $y=\frac{N-V}{N}$, and the network parameter of Eq. (\ref{normrule}) appears again. This seems to be a kind of mean value relationship, where the global average cooperation level taken over all values of the cost-to-benefit ratio also gives the local transition value $r_{0.50}$.

Finally, notice the similarity between the criterion of Eq. (\ref{normrule}) and Hamilton's rule for kin selection. Despite arising in a completely separate context, Hamilton's rule gives a genetic criterion for the emergence of altruistic behavior between individuals when the {\it genetic relatedness} of the individuals exceeds the cost-to-benefit ratio of the altruistic act. Genetic relatedness is measured by the probability that two genes randomly selected from each individual at the same locus are identical by descent \cite{Ham}. In the context of a social network, the parameter $\frac{N-V}{N}$ serves as a natural definition for a notion of {\it social relatedness}. Like genetic relatedness, social relatedness lies in the interval $\left[0,1\right)$, with larger values indicating increased relatedness. If two networks have the same fixed average agent size $V$, then there is more social cohesion in networks with larger, more influential neighbors. As a result, $N$ emerges as the parameter governing social viscosity \cite{OHLN,vBR}, where larger neighbors increasingly facilitate relatedness, and through this, cooperation. Similar parallels have been drawn before, particularly in \cite{OHLN} where a weak selection model was considered, but doesn't appear relevant to the case of strong selection considered here as it fails to distinguish between networks with different topologies, but the same average degree.

In conclusion, we have shown that a simple analytical framework, using basic ideas form the theory of complex networks, can effectively predict the success of cooperation in an evolutionary PD on varied network topologies. Moreover, the analysis suggests a network-based evolutionary rule that nicely parallels Hamilton's classical genetic rule for kin selection.


%
%
%
%

\end{document}